\documentclass[12pt]{article}

 \newread\testifexists
 \def\GetIfExists #1 {\immediate\openin\testifexists=#1
     \ifeof\testifexists\immediate\closein\testifexists\else
     \immediate\closein\testifexists\input #1\fi}

 \usepackage{gthstyle}\usepackage{amsfonts}
 \usepackage{amssymb}
 \immediate\openin\testifexists=e
     \ifeof\testifexists\immediate\closein\testifexists\else
     \immediate\closein\testifexists\input e\fipsf

 \def\Bbb#1{\setbox0=\hbox{$\tt #1$}  \copy0\kern-\wd0\kern .1em\copy0}

 \def\bbf#1{\setbox0=\hbox{$#1$} \kern-.025em\copy0\kern-\wd0
         \kern.05em\copy0\kern-\wd0 \kern-.025em\raise.0433em\box0}

 \immediate\openin\testifexists=a
     \ifeof\testifexists\immediate\closein\testifexists\else
     \immediate\closein\testifexists\input a\fimssym.def  

 \def\a{\alpha}      \def\b{\beta}   \def\g{\gamma}      
 \def\d{\delta}        \def\e{\varepsilon} 
       \def\l{\lambda}      \def\m{\mu}
             \def\vv{\varphi}    \def\n{\nu}
             \def\r{\varrho}     \def\s{\sigma}  
         \def\th{\theta}  
               
 \def\w{\omega}      \def\W{\Omega}

  \def\ra{\rightarrow}
 \def\bal{$\bullet$} 
 \def\dd{{\rm d}}

 \def\deff{\ {\buildrel{\rm def}\over{=}}\ }
 \def\iss{\ =\ }

 \def\fract#1#2{{\textstyle{#1\over#2}}}
 \def\ffract#1#2{\raise .3 em\hbox{$\scriptstyle#1$}\kern-.25em/
                 \kern-.2em\lower .2 em \hbox{$\scriptstyle#2$}}
 
 \def\half{\fract12}  
 
 \def\part#1#2{{\partial#1\over\partial#2}}

 \newcommand{\tl}[1]{\tilde{#1}}              \newcommand{\Tr}{{\mbox{Tr}}\,}
                     \newcommand{\fn}{\footnote}
 \newcommand{\nn}{\nonumber\\[2pt]}             
 \newcommand{\be}{\begin{eqnarray}}             \newcommand{\ee}{\end{eqnarray}}
 \newcommand{\bi}[1]{\begin{itemize}\item[#1]}         \newcommand{\itm}[1]{\item[#1]}
       \newcommand{\ei}{\end{itemize}}
 \newcommand{\eqn}[1]{(\ref{#1})}

 \newcommand{\crlb}[1]{\label{#1}\\[2pt]}
 \newcommand{\eela}[1]{\quad\hbox{\scriptsize{#1}}\label{#1}\end{eqnarray}}
 \newcommand{\eelb}[1]{\label{#1}\end{eqnarray}}
 
 \newcommand{\newsecb}[2]{\section{#1}\label{#2}\setcounter{equation}{0}}
 
 \newcommand{\nolabels} {\def\eel{\eelb} \def\crl{\crlb} \def\newsecl{\newsecb}}

\newcommand\publishversion{\nolabels\setlength{\textheight}{9in}\setlength{\oddsidemargin}{0in}
    \setlength{\textwidth}{6.3in}\setlength{\topmargin}{-0.1in}}

\publishversion 
\begin{document} \begin{titlepage}

\title{\normalsize \hfill ITP-UU-08/17 \\ \hfill SPIN-08/16 \\ \hfill {\tt arXiv:yymm.nnnn}\\ \vskip 20mm
\Large\bf A LOCALLY FINITE MODEL FOR GRAVITY}

\author{Gerard 't~Hooft}
\date{\normalsize Institute for Theoretical Physics \\
Utrecht University \\ and
\medskip \\ Spinoza Institute \\ Postbox 80.195 \\ 3508 TD
Utrecht, the Netherlands \smallskip \\ e-mail: \tt g.thooft@phys.uu.nl \\ internet: \tt
http://www.phys.uu.nl/\~{}thooft/}

\maketitle

\begin{quotation} \noindent {\large\bf Abstract } \medskip \\
Matter interacting classically with gravity in 3+1 dimensions usually gives rise to a continuum of
degrees of freedom, so that, in any attempt to quantize the theory, ultraviolet divergences are
nearly inevitable. Here, we investigate matter of a form that only displays a finite number of
degrees of freedom in compact sections of space-time. In finite domains, one has only exact,
analytic solutions. This is achieved by limiting ourselves to straight pieces of string, surrounded
by locally flat sections of space-time. Globally, however, the model is not finite, because
solutions tend to generate infinite fractals. The model is not (yet) quantized, but could serve as
an interesting setting for analytical approaches to classical general relativity, as well as a
possible stepping stone for quantum models. Details of its properties are explained, but some
problems remain unsolved, such as a complete description of the most violent interactions, which
can become quite complex.

\end{quotation}

\vfill \flushleft{\today}

\end{titlepage}

\eject

\newsecl{Introduction: Gravity in 2+1 dimensions}{intro}
Classical point particles, interacting only gravitationally in 2+1 dimensions, require a limited
number of physical degrees of freedom per particle.\cite{SB}\cite{AS}\cite{gthDJ84} Although they
are surrounded by locally flat space-time (if the cosmological constant is taken to be zero),
space-time may globally form a closed, compact universe.\cite{gth2+1} The classical equations can
be solved exactly, and for this reason this is a magnificent model for a complete, exactly solvable
cosmology. Many researchers are more interested in quantized theories, where the point particles
are removed as being unwanted topological defects, to be replaced by non-trivial global topological
features of the universe. Such theories however have no local degrees of freedom at all, so from a
conceptual point of view they are actually further removed from the physical world than our
gravitating point particles.

\begin{figure}[b] \setcounter{figure}{0}
\begin{quotation}
 \epsfxsize=80 mm\epsfbox{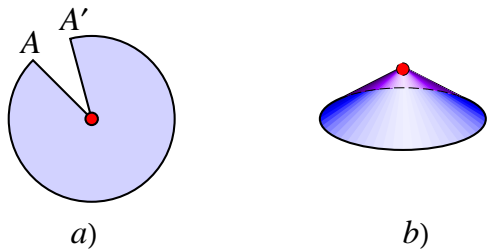}
  \caption{$a$) 2 dimensional space surrounding a point particle. Points \(A\) and \(A'\) are identified.
  $b$) artist's impression of 2 dimensional space embedded in higher dimensions.}
  \label{pp.fig}\end{quotation}
\end{figure}

\begin{figure}[h]
\begin{quotation}
 \epsfxsize=70 mm\epsfbox{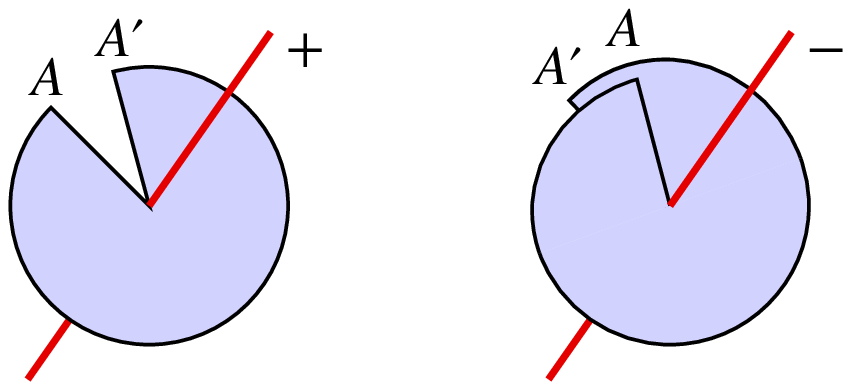}\\
 \epsfxsize=70 mm\epsfbox{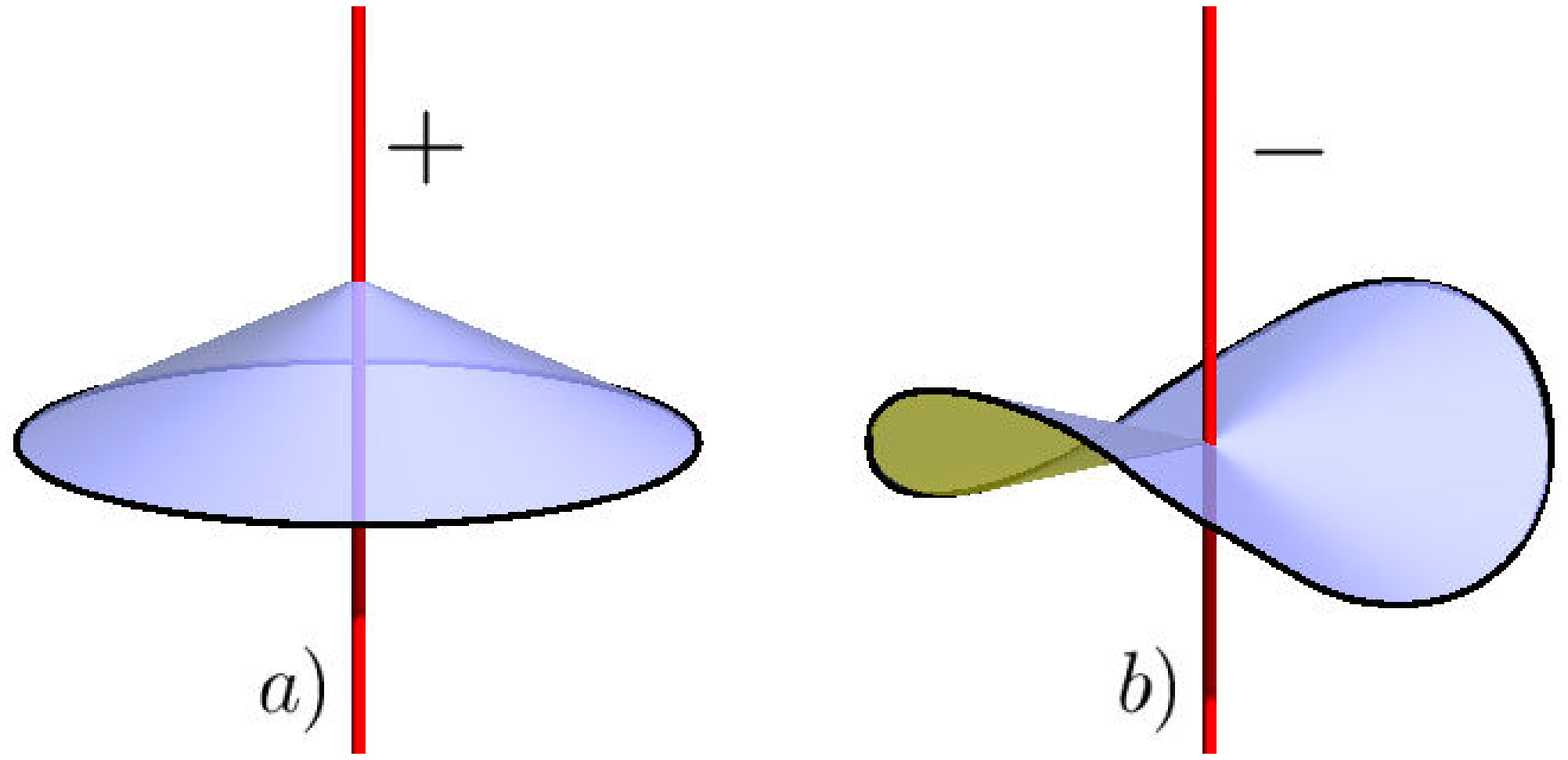}
  \caption{$a$) Cross section of 3-space surrounding a positive string, with deficit angle
  \(A\,A'\);
  $b$) Cross section of 3-space  surrounding a negative string, showing a surplus angle \(A\,A'\).}
  \label{stringwrap.fig}\end{quotation}
\end{figure}

Indeed, the importance of the model of gravitating point particles in a locally flat 2+1
dimensional space-time, is still severely underestimated. It will serve as a starting point for the
3+1 dimensional theory discussed in this paper. In 2+1 dimensions, pure gravity (gravity without
matter in some small section of space-time) has no physical degrees of freedom at all. This is
because the Riemann curvature \(R_{\a\b\g\d}\) can be rewritten in terms of a symmetric \(3\times
3\) matrix \(Q^{\m\n}\) as follows:
    \be R_{\a\b\g\d}=\e_{\a\b\m}\,\e_{\g\d\n}\,Q^{\m\n}\ , \eel{Qmunu}
so that the Ricci curvature is
    \be R_{\a\g}={R_{\a\b\g}}^\b=(g_{\m\n}g_{\a\g}-g_{\m\g}g_{\a\n})Q^{\m\n}=Q^\m_\m\,
    g_{\a\g}-Q_{\g\a}\ ;\crl{RicciQ} \hbox{so that}\qquad
    Q_{\m\n}=\half R^\a_\a \,g_{\m\n}-R_{\m\n}\ . \eel{QRicci}
Clearly, if matter is absent, \(R_{\m\n}\) vanishes, and therefore so do \(Q_{\m\n}\) and
\(R_{\a\b\g\d}\). Conversely, a point particle represents point curvature. Thus, particles are point
singularities surrounded by flat space-time. A particle at rest can be described as in
Figure~\ref{pp.fig}. The wedge is stitched closed, so that the points \(A\) and \(A'\) are identified.
The defect angle can directly be identified with the particle's rest mass. One can show that systems
containing several point particles in motion, while the total momentum is kept zero, is surrounded by a
conical space-time, of which the deficit angle can be identified with the total energy.\fn{There may
however also be a time shift when a curve is followed around the cone. This time shift is then
identified with total angular momentum.} Only particles with negative rest mass, or systems with
negative energy, are surrounded by a conical space-time with negative deficit angle, or surplus angle,
see also Fig.~\ref{stringwrap.fig}$b$.

    When such a particle is set in motion, the surrounding space-time is described by performing
Lorentz transformations upon the stationary case of Fig.~\ref{pp.fig}. One then can study systems with
many particles, by viewing space-time as a tessellation of locally flat triangles, or polygons. A number
of surprises are encountered: \bi{\bal} Using fast moving, heavy particles, a space-time can be created
that appears to allow for the existence of closed timelike curves\cite{Gott}, much like in G\"odel's
universe\cite{Godel}. This would clash with fundamental principles of causality, but one can also show
that in physically realistic models such configurations cannot occur, because a universe that contains
such a ``Gott pair" would actually collapse to a point before the timelike curve could be
closed\cite{gthDJ92}. Any closed timelike curve that one would be tempted to construct would pass
through a non-existing region of the universe. \itm{\bal} If all particles in a 2\,+\,1 dimensional
universe would be stationary or nearly stationary, the two-dimensional integrated scalar Ricci tensor,
\(\int\d^2x\,\sqrt{g}\, R\), would have to be positive, so with a sufficient number of particles the
spacelike part of this universe would always close into an \(S(2)\) geometry. Its timelike coordinate
can form a compact dimension (featuring both a \emph{bang} and a \emph{crunch}), or a semi-infinite one,
with either a bang or a crunch. \itm{\bal} With fast moving particles one can however also form 2-d
surfaces with higher genus, without requiring negative mass particles.\cite{Kadar}\ei

\noindent Quantization of this system is often carried out `as usual',\cite{WitCarl} but there are
delicate problems, having to do with the fact that we are dealing with a strictly finite universe, so
that the role of an `observer' is questionable, and the statistical interpretation of the wave function
is dubious because the finiteness of the universe prohibits infinite sequences of experiments to which a
statistical analysis would apply. Carrying quantization out with care, one first observes that
evidently, time is quantized into `Planck time' units\cite{gthtime}. This is easily derived from the
fact that the hamiltonian is an angle and it is bounded to the unit circle. Consequently, a quantum
theory cannot be formulated using differential equations in time, but rather one should use evolution
operators that bridge integral time segments. This indeed can be regarded as a first indication of some
sort of space-time discreteness, which we will encounter later in a more concrete way. However, a
confrontation with foundational aspects of quantum mechanics appears to be inevitable.

In this paper, the question is asked whether a similar ``finite" theory can also be formulated in 3\,+\,1
dimensions. This is far from obvious. One first notices that the absence of matter now no longer guarantees
local flatness, since the Ricci curvature \(R_{\m\n}\) can vanish without the total Riemann curvature \(R_{\
\b\m\n}^\a\) being zero. However, one still can decide to view space-time as a tessellation of locally flat
pieces. The defects in such a construction again may represent matter. The primary defects one finds are
direct generalizations of the 2\,+\,1 dimensional case. Take a particle-like defect in 2-space. In 3+1
dimensions, such defects manifest themselves as strings.

Our starting point is that, indeed, straight strings are surrounded by a locally flat metric. In
Section~\ref{strings}, we recapitulate the well-known derivation of the metric near such a string. Then, in
Section~\ref{motion}, moving strings are described. These topics are quite elementary but we need these
discussions to initiate the mathematical derivations that come next. In Section~\ref{connect}, joints between
strings are introduced. Then we arrive at collisions. The orthogonal case (Section~\ref{ortho}) seems to be
the easiest case, though we will see that there is a catch, at the end. The most difficult case occurs when
two strings approach at an angle. There are two possibilities. The ``quadrangle" final state is discussed in
Section~\ref{slanted}. It leads to delicate mathematical relations between \(SL(2,C)\) matrices. We needed to
do some computer algebra, the result of which was deferred to the Appendix. This algebra reveals that, at the
highest relative velocities of the approaching strings, the quadrangle final state is ruled out, so that more
complex final states are expected. An attempt to describe these is made in Section~\ref{next}.

\newsecl{Strings}{strings}

It could be that matter is always arranged in such a way that it can be regarded as defects in a locally
perfectly flat space-time. As opposed to technical approaches towards solving General Relativity\cite{Regge},
we now regard matter of this form as elementary. The novelty in this idea is that, in spite of matter being
distributed on subspaces of measure zero, we still insist that it obeys local laws of causal behavior. Let us
see how this looks.

By simply adding the third space dimension as a spectator, orthogonal to the first two, we find flat 3
dimensional space-time surrounding a string. Indeed, the energy momentum tensor of a straight, infinite,
static string pointing in the \(z\)-direction, is
 \be T_{\m\n} = t_{\m\n}\,\d^2(\tl x)\ ,\qquad t_{33}=-t_{00}=\r\ ,\qquad\hbox{all other
 }\ t_{\a\b}=0\ , \eel{stringemtensor}
where \(\r\) is the string tension parameter. When \(\r\) is small, the metric generated by such a
string is found by slightly smearing the delta peak. The curvature is only in the transverse
coordinates \(\tl x\). Choosing conveniently scaled coordinates and replacing \(\d^2(\tl x)\) by
 \be {1\over \pi \e^2}\,\th(\e-\tl r)\ , \eel{smeareddelta}
 where \(\tl r=|\tl x|\), we find that for \(\tl r\leq\e\), the transverse components of the
metric must be those of a sphere,
 \be \dd s^2=\dd \tl r^2+ \sin^2\tl r\,\dd\vv^2+\dd z^2-\dd t^2\ ,\qquad \tl r\leq\e \ . \eel{stringmetric}
At \(\tl r=\e\), the transverse metric changes into that of a cone. The cone that touches the
sphere at that point generates the metric
 \be \dd s^2=\dd r_1^2+( r_1\cos\e)^2\dd\vv^2+\dd z^2-\dd t^2\ ,\qquad r_1\cos\e\geq\sin\e \ ,
 \eel{conemetric}
where the coordinate \(r_1\) is matched to the coordinate \(\tl r\) at the point \(\tl r=\e\),
\(r_1=\tan\e\). The deficit angle \(\a\) of the cone is \(2\pi(1-\cos\e)\).

Inside the smeared region, where the metric is that of Eq.~\eqn{stringmetric}, we have the Ricci
curvature
 \be R_{xx}=R_{yy}=1\ ;\quad R_{zz}=R_{tt}=0\ , \eel{Riccicurv} while on the conical metric
\eqn{conemetric}, the curvature vanishes. Substituting Eq.~\eqn{stringemtensor} into Einstein's
equation, one then gets
 \be 8\pi G\r{1\over\pi \e^2}=1\ ,\eel{eqa}
so that for small values of \(\e\), the deficit angle can be written as
 \be \a=2\pi(1-\cos\e){8G\r\over\e^2}\iss 8\pi G\r . \eel{astring}
We quickly return to the old coordinates where the delta peak is very sharp. Space-time surrounding a
string is then seen to be as sketched in Fig.~\ref{stringwrap.fig}$a$. When the string constant \(\r\)
is large, we \emph{redefine} it to be the one generating exactly the deficit angle \(\a=8\pi G\r\). Note
that a \textit{positive} string constant leads to a \textit{deficit} angle. A negative string constant
would produce a \textit{surplus} angle, see Fig.~\ref{stringwrap.fig}$b$. Normally, in string theory, a
negative string constant would make a string highly unstable, as it contains positive pressure and
negative energy. In our case, however, strings are constrained to form straight lines, and therefore
this instability has no effect at small distances. We later may wish to include such negative strings in
our models. We leave this open for the time being, but unless indicated specifically, we will usually be
discussing strings with positive string constants and thus with positive deficit angles.

A more conventional derivation of the locally flat metric surrounding a string can be found for
instance in \cite{Betal}.

\newsecl{Moving strings}{motion} A description of a multitude of static strings is now straightforward, in
principle. However, already here, one may expect considerable complications. The system of static strings is
not unlike the 2\,+\,1 dimensional universe with moving point particles, where the time coordinate is
replaced by the coordinate \(z\) of the static string system. We know that the 2\,+\,1 dimensional world has
either a big bang singularity or a big crunch\cite{gth2+1}. Similar ``infrared" singularities might show up
in the static string system. This question will not be further pursued here. It is the local properties of
the model that we will investigate further.

Moving strings impose important questions concerning the internal consistency of the model. A moving
string can be characterized in different ways. Firstly, one can specify the orientation vector
\(\vec\w\) of the string, normalized such that its norm coincides with half the deficit angle,
\(\half\a=4\pi\,G\r\) when the string is at rest: \be |\vec\w|=\half\a\ . \eel{orientw} In addition then
we specify the velocity vector \(\vec v\) of the string. But, noticing that the string is invariant
under boosts in the direction \(\vec\w\), only the component of \(\vec v\) orthogonal to \(\vec\w\)
matters, so we limit ourselves to the case \be\vec v\cdot\vec\w=0\ . \eel{vwortho} Finally, the position
of the string at \(t=0\) should be specified. This requires another vector orthogonal to \(\vec\w\). All
in all, this requires 3\,+\,2\,+\,2\,=\,7 real parameters, of which 5 are translationally invariant, and
2 can be set to zero by a spacelike translation in 3-space.

Alternatively, we can specify the string's characteristics by giving the element of the Poincar\'e
group that describes the holonomy along a non-contractible cycle \(C\) around the string. For
static strings through the origin of 3-space, this is just the pure rotation operator, which will
be denoted as \(U(\vec\w)\). For strings moving with velocity \(\vec v\) through the origin, this
is the element \(B(\vec v)\,U(\vec\w)\,B(-\vec v)\) of the Lorentz group, where \(B(\vec v)\) is
the element of \(SL(2,C)\) that represents a pure Lorentz boost corresponding to the velocity
\(\vec v\).

If the string does not move through the origin, we get a more general element of the Poincar\'e
group.

Notice however, that an arbitrary element of the Lorentz group is specified by 6 parameters, not 5,
and the elements of the Poincar\'e group by 10 parameters, not 7. This means that not all elements
of the Poincar\'e group describe the holonomy of a string. Firstly, ignoring the translational
part, the pure Lorentz transformation \(Q\) associated to the closed curve \(C\) has to obey one
constraint:

\textit{There must be a Lorentz frame such that, in that frame, \(Q\) is a pure rotation in
3-space.}

\noindent Since we plan to describe these Lorentz transformations in terms of their representations
in \(SL(2,C)\), we identify this as a constraint on the associated \(SL(2,C)\) matrices. Write
 \be Q=B(\vec v)\,U(\vec\w)\,B(-\vec v)\ ,\eel{QBUB}
 where \(B(\vec v)\) are \(2\times 2\) matrices representing boosts with velocity \(\vec v\)
(we will see shortly that pure boosts are represented by \emph{hermitean} \(2\times 2\) matrices,
Eq.~\eqn{vsh}), and \(U\) is a unitary matrix representing a pure rotation. When \(\vec\w\) points in
the \(z\) direction, we have \be U(\vec\w)=\pmatrix{e^{i\w}&0\cr 0&e^{-i\w}}\ , \eel{wmatrix} so that
\be\Tr(U)=2\cos\w\ . \eel{TrU} Since the trace is invariant under rotations and the boosts \eqn{QBUB},
it follows quite generally that
 \be\hbox{Im}(\Tr(Q))&=&0\ ; \crl{Trreal} |\hbox{Re}(\Tr(Q))| &\le& 2\ .\eel{reTr}
Eq.~\eqn{Trreal} fixes one of the real variables of the Lorentz transformation \(Q\). Inequality
\eqn{reTr} is important in a different way. A generic \(SL(2,C)\) matrix can de written in a basis
where it is diagonal. Because the determinant is restricted to be 1, the diagonal form is then
 \be Q=\pmatrix{z&0\cr 0&1/z}\ , \eel{generalQ}
where \(z\) can be any complex number. Imposing \eqn{Trreal} leaves two options: either \(z\) is on the
unit circle -- in which case it represents a pure rotation in 3-space, or it is a positive or negative
real number. In the latter case, \(Q\) is a pure Lorentz boost, and this is when Ineq.~\eqn{reTr} is
violated. It describes the holonomy of something that is quite different from a string. We return to
that at the end of this section.

The second restriction to be imposed on the holonomy of a physical string is the translational part
of the element of the Poincar\'e group. We just saw that in the frame where \(Q\) is diagonal, the
string is static, and its position should be characterized by a vector orthogonal to \(\vec \w\).
Using the same notation \eqn{QBUB} to write the full Poincar\'e element \(P\), we add a
displacement operator \(D(\vec u)\) (\(\vec u\) being the displacement vector):
 \be P(\a^\m_{\ \n},y^\m)=B(\vec v)\,D(\vec u)\,U(\vec\w)\,D(-\vec u)\,B(-\vec v)\ ,\eel{PBDUDB}
where \(\a^\m_{\ \n}\) is the generator of the Lorentz transformation, and \(y^\m\) is the 4
dimensional displacement vector:
 \be (P\,x)^\m\deff (Q\,x)^\m+y^\m\deff L(\a)^\m_{\ \n}\,x^\n+y^\m\ . \eel{Poinc}
Henceforth, expressions such as \(Q\,x\) stand short for a \(4\times 4\) matrix \(L^\m_{\ \n}\)
acting on the 4-vector \(x^\m\). In \(SL(2,C)\) notation, this would read
 \be (Qx)^0 \mathbb{I}+(Qx)^a\,\s_a=Q(x^0\mathbb{I}+x^a\s_a)Q^\dagger\ , \eel{xtransform}
where \(\s_a\) are the three Pauli matrices. From Eq.~\eqn{PBDUDB}, we have
 \be y=B(\vec v)\,(u\,U(\vec\w)-U(\vec\w)\,u)\,B(-\vec v)\ . \eel{yu}

In the static case, \(B=\mathbb{I}\), \(y\) can neither have a time component nor a component
parallel to \(\vec\w\). The time component could be introduced to describe a spinning string,
analogous to a spinning point particle in \(2+1\) dimensional gravity, but the problem with that is
that such a space-time would possess closed timelike curves (CTC); thus, causality would be a
problem.

A component of \(y\) in the \(\vec\w\) direction could be introduced as a generalization of the
string concept; it would describe a string with torsion -- one could call that a ``spring". We will
not discuss springs further, but we have to keep this possibility in mind. Barring spin and
torsion, gives two constraints on the vector \(y^\m\).

Specifying the element of the Poincar\'e group that describes the holonomy associated to a
non-contractible curve around a string, specifies its position provided \(\vec\w\) does not vanish
(vanishing strings have no specified position). To find the location of a string if \(P\) is given
is easy: just solve the equation
 \be Px=x\ , \eel{Pposition}
In the static case, this gives \be x=s\,\vec\w+t\,e^0\ , \eel{twospace} where \(e^0\) is the unit vector
in the time direction, while \(s\) and \(t\) are free parameters. Thus, we find the string world sheet.

We end this section with a few important facts about Lorentz transformations in the \(SL(2,C)\)
representation, for future use.
 \bi{{[1]}} Pure rotations in 3-space are described by the subgroup \(SU(2)\) of
\(SL(2,C)\). Thus, \(Q\) is a pure rotation iff \(Q^\dagger=Q^{-1}\). We often write such a matrix as
\be U(\vec\w)=c\,\mathbb{I}+i\sum_as_a\s_a\ , \eel{Qunitary} where \be c=\sqrt{1-{\sum}_as_a^2}\
,\eel{cs} and \(\s_a\) are the three Pauli matrices. \(s_a\) is the rotation vector \(\vec\w\) but with
a different normalization: \be s_a={\sin|\w|\over|\w|}\w_a\ . \eel{somega} For a pure rotation along the
\(z\)-axis, see Eq.~\eqn{wmatrix}, \(c=\cos(\w)\) and \(s_3=\sin(\w)\). Note that, as before
(Eq.~\eqn{orientw}), \(\w\) is half the full rotation angle.
 \itm{{[2]}} Pure Lorentz boosts \(B(\vec v)\) are Hermitean \(SL(2,C)\) matrices. Diagonalizing such a
 matrix corresponds to rotating \(\vec v\) into the \(z\)-direction. The matrix then takes the form
 \be B(\vec v)=\pmatrix{r&0\cr 0&1/r}\ ,\qquad v={r^2-1\over r^2+1}\ . \eel{vdiag}
In general, \be B(\vec v)=ch\,\mathbb{I}+\sum_a sh_a\s_a\ , \eel{vherm} where the vector \(\vec{sh}\) is
\(\vec v\) apart from a normalization:
 \be sh_a={v_a\over \sqrt{1-v^2}}\ ;\qquad ch={1\over\sqrt{1-v^2}}\ . \eel{vsh} Notice that
 \(\Tr(B)\ge 2\). Hermitean matrices with \(\Tr(B)\le -2\) are equivalent to \(-B\) since all
 \(SL(2,C)\) matrices \(Q\) describe the same Lorentz transformation as \(-Q\).
 \itm{{[3]}} Any Lorentz transformation \(Q\) can be associated to a vector \(\vec v\) and a vector
 \(\vec\w\) such that it is the product of a pure boost and a pure rotation:
 \be Q=B(\vec v)\,U(\vec\w)\ . \eel{boostrot} Proof: define the matrix \(R\) by
  \be R=Q\,Q^\dagger\ , \eel{RQQdag} and notice that, since \(R\) is hermitean and positive definite,
it can be written as
  \be R=B^2\ , \eel{RBsquared} where \(B\) is also hermitian and, since \(\det(Q)=1\), also
 \(\det(B)=1\). Diagonalizing \(R\) gives us \(B\) in diagonal form, and its eigenvalues (whose
product is one), can be matched with the boost velocity \(\vec v\) which is again in the
 \(z\)-direction in this frame. Finally, define \(U\) as
  \be U=B^{-1}Q\ ,\qquad U\,U^\dagger=B^{-1}Q\,Q^\dagger B^{-1}=\mathbb{I}\ . \eel{UUdagger}
 \(U\) is unitary, therefore it describes a pure rotation in 3-space.
  \itm{{[4]}} Given any Lorentz transformation \(Q\) with \(\hbox{Im}(\Tr(Q))=0\), then there exists
a pure boost operator \(B(\vec v)\) and either a rotation \(U(\vec\w)\) or another pure boost
\(B(\vec{v'})\) such that
  \be \hbox{either}\quad Q=B(\vec v)\,U(\vec\w)\,B(-\vec v)\ ,\qquad\hbox{or}\quad Q=\pm B(\vec v)\,B(\vec
{v'})\,B(-\vec v)\ . \eel{Qboosts}
 Proof: looking at the eigenvalues of \(Q\), one finds that, since their product is one and the sum
is real, they either match the eigenvalues of \(U(\vec\omega)\) or those of \(\pm B(\vec{v'})\). A
matrix that diagonalizes \(Q\) can be written as \(B(\vec v)U(\vec\w_1)\) due to the previous
theorem. So we have
   \be Q=B(\vec v)\,U(\vec\w_1)\,R\,U^{-1}(\vec\w_1)B^{-1}(\vec v)\ , \eel{Qdiag} where \(R\) stands
for the matrix that is either a rotation or a boost. This rotation or boost was in the
\(z\)-direction, since \(R\) was diagonal. The matrix \(U(\vec\w_1)\) rotates that vector into any
other direction in 3-space.\ei

\noindent We now return to Eqs.~\eqn{Pposition} and \eqn{twospace} for the string world sheet.
Suppose we have a \(Q\) that obeys Eq.~\eqn{Trreal} but not the inequality~\eqn{reTr}. Then, in
some Lorentz frame, this is a pure boost rather than a pure rotation. Write it as
 \be Q=\pmatrix{r&0\cr 0&1/r}\ , \eel{Qboost} where we took the boost to be in the \(z\)-direction. The
 equation for the ``world sheet", \(Qx=x\), now leads to \be z=0\ , \quad t=0\ , \eel{boostsheet}
 in other words, the transverse plane at \(t=0\). This is a spacelike surface rather than a
 timelike string world sheet. What we find here is a ``tachyonic" string. Such elements will
 be difficult to incorporate in a viable gravity model, if we wish to have some version of
 causality. We will therefore attempt to avoid structures for which the holonomy is a boost.

\newsecl{Connecting strings}{connect}

\begin{figure}[h]
\begin{quotation}
 \epsfxsize=80 mm\epsfbox{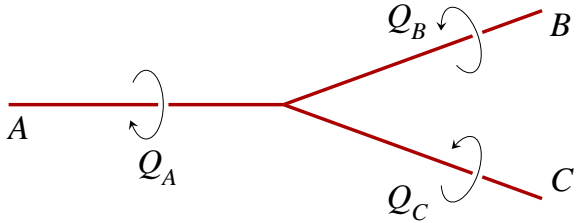}
  \caption{junction connecting three strings. The holonomy \(C_A\) of string \(A\) equals the product
  \(C_B\,C_C\) of the holonomies of strings \(B\) and \(C\)}
  \label{junction.fig}\end{quotation}
\end{figure}

We could try to limit ourselves to having only infinite, straight strings in our model, but as soon as
collisions are considered -- and we will argue that these are inevitable -- one must face the presence
of strings with finite lengths. These strings must then be connected to other strings in junctions, see
Fig.~\ref{junction.fig}. Just as our infinite strings, junctions are also surrounded by flat space
\cite{Betal}. The rules for connecting three strings \(A\), \(B\) and \(C\) are as follows.
 \bi {$i$)} The junction at time \(t\) must lie on a point \(x(t)\) that is on the world sheet
of the three strings and is a straight line. Thus, the three world sheets must have a straight line
in 4-space in common. This line is a solution of
  \be Q_Ax=Q_Bx=Q_Cx=x \eel{junctioneq} (which is, again, in the 4 dimensional notation).
Depending on whether this line is timelike, spacelike or lightlike, there are three classes of
junctions: \emph{subluminal}, \emph{superluminal} and \emph{lightlike}. Superluminal junctions will
be seen to come in two types.

 \itm{$ii$)} The fact that the surrounding space-time is flat implies that the holonomies must
 match: \( Q_A=Q_BQ_C\) or, if we take all strings pointing towards the junction (so that \(Q_A\)
 turns into its inverse),
  \be Q_AQ_BQ_C=\mathbb{I}\ . \eel{holonomymatch} \ei
In the case of a superluminal junction, the three strings have a spacelike line in
common. This means that a special Lorentz frame exists where this is a straight line in
the \(z\)-direction that is instantaneous in time. The three connecting strings are
parallel in that frame, but they may have different velocities. We either have one string
splitting in two, or two strings merging into one. The first of these cases is impossible
to reconcile with local causality, but the latter, in principle, is: two parallel strings
meet and subsequently merge. Since the time reverse of this event violates causality,
this would be an example of \emph{information loss}. At first we will find that this kind
of events may be difficult to avoid, but we will show how this can nevertheless be
achieved if we so wish.

In more general Lorentz frames, superluminal junctions can be easily recognized as they
describe a pair of strings opening up or closing like a superluminal zipper. A
superluminal zipper that is opening up will have to be avoided at all times; the closing
(joining) superluminal zipper is a curious case of information loss.

Lightlike and timelike (subluminal) junctions are fine.

If we wish two strings \(A\) and \(B\) to meet at one subluminal junction for an extended amount of
time, then this gives three restrictions on the associated holonomies \(Q_A\) and \(Q_B\) alone:
first, the product of their holonomies must again be a string holonomy, or
 \be \hbox{Im}(\Tr(Q_A\,Q_B))=0\ ,\qquad |\hbox{Re}(\Tr(Q_A\,Q_B))|\leq 2 \ . \eel{stringsmeet}
From this, one can show that in a Lorentz frame where string~\(A\) is static and pointing in the
\(z\)-direction, we have
 \be Q_A=\pmatrix{e^{i\w}&0\cr 0&e^{-i\w}}\ ,\qquad
 Q_B=\pmatrix{a_1+ia_2&b_1+ib_2\cr\m(-b_1+ib_2)& a_1-ia_2}\ , \eel{subljunction}
where all coefficients are real. Our second restriction now is that, in Eq.~\eqn{subljunction},
 \be \m>0\ , \eel{mupos}
which corresponds to a subluminal junction. If \(\m<0\) we have a superluminal junction. If \(\m=1\) the
string \(B\) is static as well. One easily checks that then \(Q_B\) is unitary and hence a pure
rotation. The case for general positive \(\m\) is obtained by Lorentz boosting in the only allowed
direction, the \(z\)-direction (otherwise, \(A\) would not remain static). Note that such a boost is
described by Eq.~\eqn{Qboost}.

Finally, of course, the displacement vectors of the Poincar\'e group elements must also match.

\begin{figure}[h]
\begin{quotation}
 \epsfxsize=80 mm\epsfbox{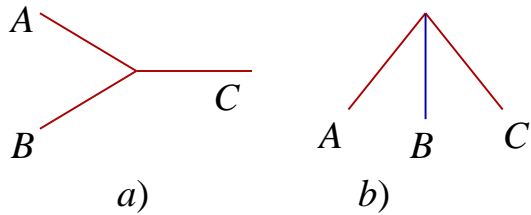}
  \caption{Two types of string junctions. In case $a$, there is always an obtuse angle present.
This is the case when all three strings are positive or all three are negative. In case $b$, one of the
strings, in fact string \#$B$, has a sign opposite to the sign that the other two have in common.}
  \label{junctions.fig}\end{quotation}
\end{figure}

Thus, we will be specially interested in the case where, for every string junction, there exists a
Lorentz frame where all three strings are static. If the string constants are large, so that the deficit
angles (or possible surplus angles) are large, the situation is a bit complicated, since at a junction
the three strings appear not to lie in a single plane. If the string constants ar weak, one discovers
that, in principle, there are two types of subluminal junctions. They are sketched in
Figure~\ref{junctions.fig}. In the first case, see Fig.~\ref{junctions.fig}$a)$, either all deficit
angles are positive or they are all negative (i.e., all surplus angles are positive). This we will refer
to as a \emph{regular} junction. The strings behave as elastic bands connected at a point: each string
appears to pull the two others towards it. In the case one string has a sign opposite to the two others,
one gets the situation sketched in Fig.~\ref{junctions.fig}$b)$: it is the situation that can be deduced
from the previous case by replacing the one string with the exceptional sign by an opposite-sign string
pointing in the opposite direction.

\begin{figure}[h]
\begin{quotation}
 \epsfxsize=40 mm\epsfbox{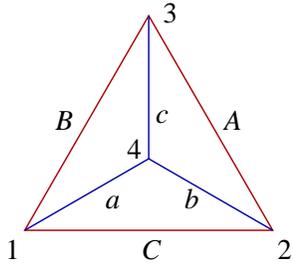}
  \caption{Six string segments connected into a triangle, forming a `localized' particle.}
  \label{localstr.fig}\end{quotation}
\end{figure}

This information is useful if one wants to investigate whether constructions can be made with only
finite extensions in space. Figure~\ref{localstr.fig} shows an example of this. We have three strings,
\(A,\ B\) and \(C\) forming a triangle, and three others, \(a,\ b\) and \(c\) that connect the three
points to a point in the middle. The junctions 1, 2 and 3 are irregular because they contain only sharp
angles. Clearly, the strings \(a,\ b\) and \(c\) must have signs opposite to the signs of \(A,\ B\) and
\(C\). Junction number 4 is a regular one.

In general, it is easy to argue that finite size constructions with only positive sign strings
cannot be possible, since the entire thing is surrounded by flat space; hence there is no
gravitational field; the total energy must be zero. This will not be possible with positive energy
strings.

Much of the above remains true when the string constants, apart from their signs, are large, but
things then are a bit more difficult to visualize, since space and space-time are locally but not
globally flat.

In this paper we will not attempt to completely avoid the emergence of negative string constants. This
would lead to negative energy states. One could think of addressing these at a later stage in a quantum
theory by some kind of second quantization. As long as we restrict ourselves to local behavior this
might not be a disaster, but of course the question of positive and negative string constants (deficit
angles) will have to be addressed. We will advocate to avoid superluminal junctions of the splitting
type at any stage, as these are difficult to reconcile with causality. Avoiding superluminal junctions
of the joining type will be a bit harder, but we will finally find a procedure to avoid those together
with the variety that opens up. Also all strings that violate the inequality \eqn{reTr} must be avoided
since they too are impossible to reconcile with causality. These two demands will require so much of our
attention that we will not further dwell on the signs of the string constants.

\newsecl{Orthogonal collisions}{ortho}

When we were dealing with point particles, in the 2\,+\,1 dimensional case, we could safely assume
that the particles will never collide head-on. In general, they will miss one another, and
consequently no further dynamical rules are needed to determine how an \(N\) particle system will
evolve. This will not be true in higher dimensional spaces\fn{Strings will in general not collide
head-on in a space-time of more than 4 dimensions. However, the generalizations of the objects we
discuss in this paper, in higher dimensions will be branes, not strings.} Strings in 3\,+\,1
dimensional space-time will in general not be able to avoid one another. They will cross, and in
doing so, two straight string sections will not be straight anymore after the collision.

Consider an initial state in which two strings are heading towards one another. We can always work in a
Lorentz frame where one of the strings, call it \(A\), is at rest. The conical 3-space surrounding it
has a deficit angle \(\a=2\w_A\). In the generic case, in this Lorentz frame, the second string does not
have to be oriented orthogonally to the first one. Its string constant, \(\b\), does not have to be the
same as \(\a\). Consider now the velocity vector \(\vec v\) of the second string. If it is not
orthogonal to the string \(A\), we perform a Lorentz boost in that direction. String \(A\) will stay at
rest. If \(\vec v\) is not orthogonal to the string \(B\), we replace it by one that is orthogonal to
\(B\). This way, one convinces oneself that, in general, we can limit ourselves to the case where \(\vec
v\) is orthogonal to both \(A\) and \(B\).

\begin{figure}[h]
\begin{quotation}
 \epsfxsize=110 mm\epsfbox{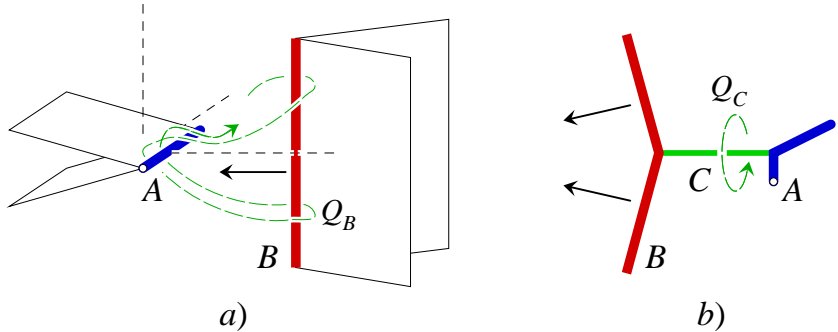}
  \caption{Orthogonal strings scattering. $a$) Initial state. String \(B\) moves towards string
  \(A\) (arrow).
  The cusps caused by their deficit angles are shown. Dashed lines are an orthogonal frame shown
  for reference. $b$) After the scattering, a new string \(C\) connecting the first two emerges.
  Strings \(A\) and \(B\) now both show a kink. The cusps in the last situation are not shown.
  The oriented dashed curves explain Eq.~\eqn{cholonomy}.}
  \label{orthosc.fig}\end{quotation}
\end{figure}

However, it is a physical limitation if we also assume \(A\) to be orthogonal to \(B\).
Just because it is special, we consider this case first. The collision event is sketched
in Fig.~\ref{orthosc.fig}. In \ref{orthosc.fig}$a$, the two strings approach one another.
They both drag a space-time cusp with them. Now, what happens when \(B\) hits \(A\), is
best understood by drawing the cusp of \(A\) in the opposite direction. The result of
that, however, is that string \(B\) is seen to have a kink. The same thing happens to
string~\(A\) itself; it develops a kink due to the cusp of \(B\). After the passage, the
two kinks must be connected by a new string, \(C\) that stretches with \(A\) and \(B\)
now moving away from one another.

Indeed, we see that, in general, the holonomy of string \(C\) is non-trivial; it is obtained from
the holonomies \(Q_A\) and \(Q_B\) of strings \(A\) and \(B\) as follows (depending on sign
conventions for \(Q_A\), \(Q_B\) and \(Q_C\)):
 \be Q_C=Q_B^{-1}\,Q_A^{-1}\,\,Q_B\,Q_A\ . \eel{cholonomy} \(Q_A\) and \(Q_B\) do not commute
because they represent rotations along two different axes. Clearly, upon crossing, two strings
produce a third stretching between them. This is why our model should not be thought of as being
globally finite. Every crossing produces more new string segments, so that, in the absence of
possible quantum effects, any regular but non-trivial initial condition will eventually create
states in which myriads of tiny string segments cover all of space-time. If the original defect
angles were relatively small, their commutators will be again much tinier, so the newly created
strings are very weak ones. Locally, however, we still have straight string segments surrounded by
flat space-time.

There is an important remark to be made here. If the original strings have been approaching each other
with velocities close to that of light then the orthogonal velocities will also be close to that of
light after the collision. However, then we can easily run into the situation that the newly produced
junctions will go faster than light: they will be superluminal. Since they will be of the ``joining"
variety, these junctions will not violate causality.

Our strategy will be to search for models where the total set of possible string holon\-omies is a
finite one, or else at least discrete, but this we leave for later invest\-igations. There is a more
urgent problem that we have to face first.

\newsecl{Slanted collisions}{slanted}

In the previous section the result was explained of a collision between two strings and a relative
velocity vector that are all orthogonal. What happens when the angles have different values?
\begin{figure}[h]
\begin{quotation}
 \epsfxsize=55 mm\epsfbox{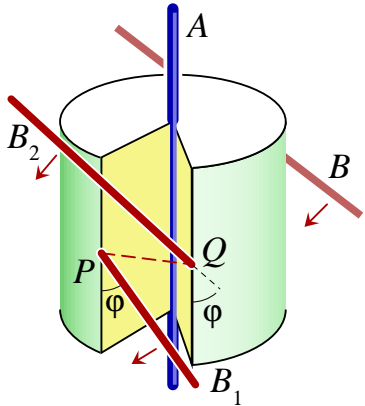}
  \caption{Scattering at an angle, \(\vv\ne 90^\circ\). The cusp of string \(A\) is shown, and the
  effect it has on string \(B\). After the collision, if the cusp would be kept
  open, these two parts would still form a straight line. Closing the cusp would move point \(P\)
  to point \(Q\). If at \(P\) the string touches the cylinder surrounding \(A\), it should do this
  now at point \(Q\). We see that then \(B_1\) and \(B_2\) do not intersect.}
  \label{slanted.fig}\end{quotation}\end{figure}

In this case, one can convince oneself that no solution is possible with a single string stretching
between the outgoing strings. This can be understood by studying the geometry, as sketched in
Fig.~\ref{slanted.fig}. But we can also verify that, in general, the holonomy~\eqn{cholonomy} is
not of the string type: it violates Eq.~\eqn{Trreal}.

\begin{figure}[h]\begin{quotation}
\epsfxsize=85 mm\epsfbox{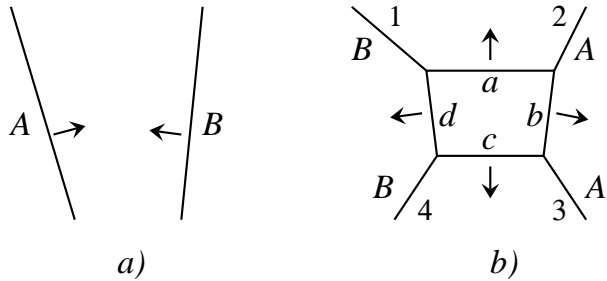}
  \caption{Scattering at an angle produces two new strands. $a$)  \(A\) and \(B\) enter. $b$)  Since
  \(A\) passes the cusp of \(B\) and \textit{vice versa}, strands \(A_2\) and \(A_3\) form angles.
  These four pieces are labelled 1---4. Four new string pieces must be further specified, here labelled
 $a$---$d$.}\label{strands.fig}\end{quotation}\end{figure}

To save the model, one can now propose the following. When two strings \(A\) and \(B\) collide at
an angle \(\vv\ne 90^\circ\), not one but two new strings appear\fn{Later, in Section \ref{conc},
we will see that even more than two new strings may emerge.}, both stretching from \(A\) to \(B\).
A single string cannot be associated with a holonomy of the form \eqn{cholonomy}, but a pair of
strings can. The question is now, whether the data of this pair of strings would be uniquely
determined by the initial characteristics of \(A\) and \(B\). To investigate this question, the
author combined analytical arguments with computer calculations, just to see how things will work
out. The topology is defined in Fig.~\ref{strands.fig}.

At first sight, it seems that we have considerable freedom to define the orientations, strengths
and velocities of the `internal' strings $a$---$d$. However, if we fix one of these, all others are
determined since the holonomies at a junction must obey Eq.~\eqn{holonomymatch}. In addition, the
strings must be properly attached to one another. As it turns out, the matching of the strings is
guaranteed if Eq.~\eqn{holonomymatch} is obeyed at all junctions, and if in addition the
\emph{string conditions}~\eqn{Trreal} and \eqn{reTr} are obeyed by all four new holonomies,
\(a\)---\(d\).

The holonomy matrices at the `external lines' 1---4 are fixed by the initial conditions.
Originally, we had, in one conveniently chosen Lorentz frame, \(Q_1=Q_B\) and \(Q_3=Q_A^{-1}\).
Here, the inverse sign arises if we decide to consider the holonomies with respect to observers
looking towards the interaction point, and the holonomy curves \(C\) are chosen to go clockwise.
Then,
 \be Q_4&=&Q_3^{-1}\,Q_1^{-1}\,Q_3\ ;\nn Q_2&=&Q_1^{-1}\,Q_3^{-1}\,Q_1\ . \eel{exthol} The last
 equation is actually the one required for consistency with the demand that
  \be Q_1\,Q_2\,Q_3\,Q_4\iss \mathbb{I}\ . \eel{totext}
Given the holonomy \(Q_a\) of the string section \(a\), the others can be defined as follows:
  \be Q_b=Q_a\,Q_2\ ;\quad Q_c=Q_b\,Q_3\ ;\quad Q_d=Q_a\,Q_1^{-1}\ ,\eel{inthol}
which is the most systematic definition, and consistency with Eq.~\eqn{totext} is ensured. We
emphasize that there is always some ambiguity in defining the Lorentz frames for the holonomies
\(Q_1\)---\(Q_4\) and \(Q_a\)---\(Q_d\), so we use Eqs.~\eqn{exthol}---\eqn{inthol} also to specify
these frames.

Let now all external holonomies \(Q_1\)---\(Q_4\) be given. How much freedom is there for
\(Q_a\)---\(Q_d\)? We do not need to consider the translation parameters in the
Poincar\'e group; these will be taken care of automatically, since there is a point
(0,0,0,0) where the colliding strings \(A\) and \(B\) first met. This point can be kept
at the origin of our coordinate frame. Thus, we consider the equations for the elements
of the Lorentz group, which are most conveniently described as \(SL(2,C)\) matrices. Each
element of the Lorentz group is characterized by 6 real variables: a rotation vector and
a velocity vector, or alternatively the four complex numbers in a \(2\times 2\) matrix
\(Q\), subject to the constraint that the complex number \(\det(Q)\) should be set equal
to 1.

The junction equations~\eqn{exthol}---\eqn{inthol} leave us the freedom to choose \(Q_a\). This
gives a space with 6 real parameters. Then the string equation~\eqn{Trreal} for \(Q_a\)---\(Q_d\),
together gives us 4 real constraints. The surviving 2-dimensional manifold is then further
constrained by the demands~\eqn{reTr}. Thus, the manifold of all possibilities is a two-dimensional
space.

To obtain somewhat more understanding of this manifold, let us consider the 8 dimens\-ional set of all
\(L(2,C)\) matrices for \(Q_a\), with\-out the non\-linear constraint concerning the determinant. The
conditions~\eqn{Trreal}, Im\(\Tr(Q_{a,b,c,d})=0\), in combination with the junction
equations~\eqn{inthol}, are 4 linear equations for the matrix elements of \(Q_a\). This leaves us with a
linear \(8-4=4\) dimensional space. Then we have the inequalities~\eqn{reTr}, which for the matrix
\(Q_a\) imply that
 \be Q_a=\pmatrix{a_1+ia_2&b_1+ib_2\cr c_1+ic_2&d_1-ia_2}\ ;\qquad |a_1+d_1|\le 2\ , \eel{reTra}
and similarly for the three other internal holonomies. Realizing that, in our 4 dimensional space,
these conditions can be written as
 \be |e_i\cdot x|\le 2\ ,\qquad i=1,\,\dots,4\ , \eel{ineqfour}
and assuming that, in general, the four vectors \(e_i\) will be independent, we see that, in the
generic case, the surviving space is a compact one: a four dimensional hypercube. We can be sure
that the inequalities~\eqn{ineqfour} give us a \emph{non empty} four dimensional space.

Next, however, we have the two constraints \be \hbox{Re}(\det(Q_a))=1\ ,\qquad
\hbox{Im}(\det(Q_a))=0\ . \eel{ReImdet} These two equations for \(Q_a\) ensure that the same
equation will hold for \(Q_b\)---\(Q_d\), because the determinant is preserved, and because
det(\(Q_i)=1\) also for the external \(Q_1\)---\(Q_4\). Now these are quadratic equations for the
coefficients of \(Q_a\), so the question whether these two equations are compatible with the
inequalities \eqn{ineqfour} and with one another is a more delicate one. It can be simplified in
the following way.

First, we can sit in a frame where string \(B\) is stationary and oriented in the \(z\)-direction,
or more precisely,
 \be Q_B=Q_1=\pmatrix{e^{i\w}&0\cr 0&e^{-i\w}}\ . \eel{Bdiag}
In that case, the condition Im\((\Tr(Q_d))=0\), \(Q_d=Q_a\,Q_1^{-1}\), see Eqs.~\eqn{inthol} and
\eqn{Trreal}, implies that the coefficients for \(Q_a\) in Eq.~\eqn{reTra} obey
 \be \hbox{Im}\left(e^{-i\w} (a_1+ia_2)+e^{i\w}(d_1-ia_2)\right)=0\quad\ra\quad d_1=a_1\ . \eel{disa}
Therefore, we can write
 \be Q_a=\pmatrix{a&b\cr c&a^*}\ , \eel{Qabc} where \(a\), \(b\) and \(c\) are complex numbers. The
 condition that det(\(Q_a\)) is real can now be written as
  \be c=-\m_1\,b^*\ , \eel{cbstar}
where \(\m_1\) is a real parameter. This is Eq.~\eqn{subljunction}. We will usually limit ourselves
to the case \(\m_1>0\), the junction with \(Q_1\) is then subluminal. The condition that the real
part of the determinant is 1 can now be written as follows:
 \be  Q_a=\l\pmatrix{1+ia_2&b_1+ib_2\cr \m_1(-b_1+ib_2)&1-ia_2}&,\quad&\m_1,\,a_2,\,b_1,\,b_2\
 \hbox{real}\ . \crl{Qaparam} \l=1/\sqrt{1+a_2^2+\m_1(b_1^2+b_2^2)}\ . &&\eel{labdacond} Note that
 choosing \(\m_1>0\) ensures that the square root is real.

The condition that the traces of \(Q_b\) and \(Q_c\) are real form two \emph{linear} conditions on
the three coefficients \(a_2\), \(b_1\) and \(b_2\) (where only the parameter \(\m_1\) appears non
linearly). Suppose that these are used to fix \(b_1\) and \(b_2\). Then we are left with \(\m_1\)
and \(a_2\) as two independent free parameters.

The question that remains is whether we can also obey the inequalities \eqn{reTr} for the string
holonomies \(Q_a\)---\(Q_d\). Those for \(Q_a\) and \(Q_d\) can easily be read off:
 \be Q_a\ :\qquad& |\l|\le 1\quad\ra\quad a_2^2\ \ \ge& -\m_1(b_1^2+b_2^2)\ ;\cr
 Q_d\ :\qquad& (\sin\w-a_2\cos\w)^2\ \ \ge& -\m_1(b_1^2+b_2^2)\ , \eel{abinequ}
which is ensured if we choose \(\m_1>0\).

Next, we can perform the same trick at the junction with \(Q_2\) or at the junction \(Q_3\). If we
choose \(Q_2\) then only one further linear constraint on the coefficients \(a_i\) and \(b_i\) follows,
and so we have two freely adjustable parameters \(\m_1\) and \(\m_2\) that now parameterize our
two-dimensional manifold. We may freely limit ourselves to positive values of \(\m_1\) and \(\m_2\) so
that we can be sure that the junctions connecting \(Q_1\) and \(Q_2\) to the quadrangle are both
subluminal. Unfortunately however, this gives us no guarantee that \(Q_3\) and \(Q_4\) will be
subluminal as well, and the line joining them, string segment \(Q_c\), is then not guaranteed to obey
the necessary inequality \eqn{reTr} that would ensure it to be a subluminal string.

There is a smarter way to proceed: we pick two opposite junctions, say \(Q_1\) and \(Q_3\). Now,
however, our numerical calculations show us a surprise. At least this author had not expected the
special thing that happens.

Suppose we first go to the Lorentz frame where \(Q_a\) and \(Q_d\) are static. Choose \(Q_1\) to be a
rotation along in the \(z\)-axis. Then \(Q_a\) and \(Q_d\) are both described by the parametrization of
Eq.~\eqn{Qaparam}, with freely adjustable \(\m_1\). The coefficients in this frame obey four real
constraints, but since the determinant is known to be real, these are actually just three new, linear
constraints.

Now perform the Lorentz transformation that makes \(Q_3\) a static rotation along the
\(z\)-axis. Again assume a freely adjustable parameter \(\m_3\) and four constraints on
the coefficients, of which only three are independent because of the determinant. One
would have thought to end up with all coefficients fixed, apart from the two freely
adjustable parameters \(\m_1\) and \(\m_3\).

But this is not what happens. The four linear constraints on the parameters are
\emph{independent}, while, instead, the two parameters \(\m_1\) and \(\m_3\) are not
independent. They are found always to be related by an equation of the form
 \be \m_3={A+B\,\m_1\over C+D\,\m_1}\ , \eel{muonemuthree}
where the coefficients \(A\), \(B\), \(C\) and \(D\) depend in a complicated way on the data that
describe the holonomies of the external lines only. \emph{This is true whenever the holonomies \(Q_1\)
and \(Q_3\) obey the string equation \eqn{Trreal}}.

This puts our problem in a different perspective: we can only succeed in devising an acceptable pattern
of a single quadrangular string loop if the coefficients \(A\), \(B\), \(C\) and \(D\) allow for
positive values for both \(\m_1\) and \(\m_3\). Conversely, if we have such a solution then we are
guaranteed that all four internal strings are Lorentz transformations of static ones, and hence they all
obey the inequality \eqn{reTr}. However, we found that the coefficients can obtain all sorts of values.
It is possible that \(A\) and \(B\) are negative while \(C\) and \(D\) are positive. In that case, it
may well be that no acceptable solution exists. We return to this case in the next section.

Since \(\m_1\) and \(\m_3\) are not independent, they only fix one parameter of our two dimensional
manifold. We can now return to introducing \(\m_2\) as the other parameter. The relation between
\(\m_2\) and \(\m_4\) is similar to the one between \(\m_1\) and \(\m_3\). So, again, we have four
coefficients \(A\), \(B\), \(C\) and \(D\), of which we must check whether they allow two positive
values for \(\m_2\) and \(\m_4\). If so, we have a solution with only subluminal junctions and
subluminal strings.

The explicit expressions for the four coefficients are too lengthy to be displayed here.
We checked numerically that indeed \(\m_1\) and \(\m_2\) are independent, so together
they can be used to search a suitable point of our two-parameter space. The resulting
relations between the coefficients \(a_i\) and \(b_i\) now completely determine their
values.

We checked explicitly with numerical examples that the above procedure appears to work flawlessly. If
the two sets of coefficients \(A\) --- \(D\) allow for positive \(\m\) values, (\(\m_1,\,\dots,\,\m_4
>0\)) this guarantees that all internal lines obey the string equation \eqn{Trreal}, that the four junctions
at \(Q_1\) --- \(Q_4\) are all subluminal, and that the four strings \(Q_a\) --- \(Q_d\) also obey the
string inequality \eqn{reTr}.

However, at each pair of antipodal junctions we have to check explicitly the existence of two positive
\(\m\) values. In Eq.~\eqn{muonemuthree}, with \(C\) normalized to one, two positive (or vanishing)
values are excluded only if\fn{The case where one or more of these coefficients are equal to zero might
be admissible, since lightlike joints do not seem to violate causality.}
 \be A < 0\ ,\quad B < 0\ ,\quad C=1\ ,\quad D>0\ . \eel{badvalues}
Thus, we have to exclude this domain for the two sets of antipodal points. It was found
however, that this domain can actually easily be entered, when the external holonomy
operators \(Q_1\) --- \(Q_4\) are far from the identity. So, if that happens, we have no
one-string-loop solution with the given topology.

Note however, that we can also choose the crossed diagrams. As in the Feynman diagrams of
quantum field theory, we have besides the original loop two crossed diagrams, such as the
one obtained by interchanging the points 2 and 3. Each of these can be tried, but still
there is no guarantee that a solution of this form will always exist.

Finally, there is another important question to ask: will the internal holonomies \(Q_a\)
--- \(Q_d\) all describe string sections with \emph{positive} string constants (positive
defect angles)? To check this is technically awkward. It means that negative energy
strings are not excluded for the time being. They are not as harmful as the strings that
violate causality, but still, one might prefer to have only states with positive local
energy densities. It seems that the wrong sign can easily come up. We decide to postpone
this question.

In fact, the orthogonal scattering case, described in Section~\ref{ortho}, would generate superluminal
junctions unless we replace the solution by our double string diagram. Here however, superluminal
junctions seem to be impossible to avoid unless we allow some of the internal string sections to have
the wrong sign for their string constants. The sign problem, therefore, appears to be difficult to
avoid.

\newsecl{Other transitions.}{next}

In the previous section it was found that there are two regions defined by the inequalities
\eqn{badvalues} (one for each diagonal), that we have to stay out of. The regions are exclusively
defined by the external holonomy matrices \(Q_1\) --- \(Q_4\), that is, by the initial string
configuration. So if we enter any one of these regions, the result of this collision cannot be the
configuration sketched in Fig.~\ref{slanted.fig}. Therefore, in that case, we have to try something
else. To de this, we made a further study of the coefficients \(A\) --- \(D\). It was found that they
enforce \(\m_1\approx\m_3\approx 1\) when the relative \emph{velocities} of the external strings are all
non-relativistic. This is the allowed region. What if strings collide relativistically?

\begin{figure}[h]
\begin{quotation}
 \epsfxsize=85 mm\epsfbox{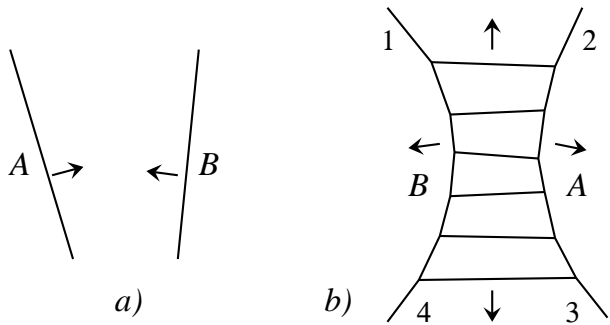}
  \caption{Scattering producing a multitude of new string sections, investigated in the text.}
  \label{multiple.fig}\end{quotation}\end{figure}

We checked the case where \(Q_1\) and \(Q_2\) are large, with possibly relativistic relative velocities,
but
 \be E\deff Q_2\,Q_3 = (Q_4\,Q_1)^{-1} \approx\mathbb{I}\ . \eel{nearlyequal}
If this case could be handled, then we can try more complicated scattering diagrams, of the kind
depicted in Fig.~\ref{multiple.fig}. If we consider a sufficiently large number of intermediate strands
in this collision process, condition \eqn{nearlyequal} can be realized in all subsegments. This would
guarantee the possibility of this multiple strand final state if the state with just two strands would
be forbidden. Even this, however, is difficult to prove. If the external holonomies obey
Eq.~\eqn{nearlyequal} then the coefficients \(A\) --- \(D\) linking diagonally opposite junctions depend
entirely on the details of the matrix elements of \(E\), regardless how close this matrix is to the
identity, as explicit algebraic calculations show. From this it follows that both the allowed and the
forbidden domains touch the point \(E=\mathbb{I}\). We were not yet able to prove that configurations
with either a single internal quadrangle or multiple strands suffice to cover all eventualities, as the
space of all possible external holonomies \(Q_1\) --- \(Q_4\) is very large.

Strings crossing over is not the only kind of ``events" that can take place in this model. We can also
encounter the situation where a string bit, of the kind that results from collisions of the type
described in the above, is reduced to zero length. It is bounded by two other junctions that herewith
merge into one. Our first try should be whether the result could again be a one-strand, two-strand, or
multiple strand final state, just like the ones described earlier. However, if we allow ourselves
strings with negative string constants, then there is a simpler final state: the one where the original
string gets a ``negative length". This is really a string where the deficit angle has switched sign.
Once it was decided to allow their presence, we could allow them here as well.

\newsecl{Discussion and conclusions}{conc}

We conclude that it may well be possible to construct a complete model for classical (\emph{i.e.}
unquantized) General Relativity with matter, which allows for the construction of piecewise exact
solutions in space-time. The model consists exclusively of piecewise straight string segments,
surrounded by locally flat regions of Minkowski space-time. This means that these string segments
actually also encompass the gravitational degrees of freedom. Interaction occurs when two pieces of
string intersect, or when the length of one or more string pieces shrinks to zero. At every
intersection, at least three (in the case of orthogonal scattering), but nearly always at least six new
string sections appear (described by the four finite segments in the quadrangle of
Fig.~\ref{strands.fig}, and remembering that the two original strings each split in two). In the latter
case, the properties of these new string segments, their string constants, as well as their orientations
and velocities, are all described by a point in a compact two parameter space. These freely adjustable
points at every interaction junction in 4-space correspond to the freedom one has in choosing the matter
interactions. This is not obviously an Euler-Lagrange system, since it cannot be mapped onto its
time-reverse. Indeed, one expects that, as time evolves, the string segments become smaller and smaller,
and more numerous as well. Also, the two-dimensional parameter space at each intersection is too small
to allow us to choose the 4 new string constants from a discrete set of a priori possibilities, and
therefore, the string constant parameter space will form a continuum, unlike what one would expect in a
realistic model of the real world.

The above are enough reasons why we will \emph{not} advocate ``quantiz\-ation" of this model along the
usual procedures. Quantization will have to go by means of the ``pre-quantiz\-ation" procedure proposed
earlier\cite{GtHDeterm}. This however will require some further refinements that will be explained in a
separate paper. The reason why we keep the subject of quantization separate is that it requires
basically new assumptions, and that the model described here could be used for different purposes.

There are quite a few open questions apart from quantization. First of all, one would like the model to
be complete, that is, give a well formulated prescription under all circumstances how the evolution
evolves. We found that many but not all pairs of strings, upon intersecting, can evolve exactly as shown
in Fig.~\ref{strands.fig}. When the relative velocities upon impact are high and the string constants
are large, more than four new strands may have to appear. One might even suspect an instability such as
the formation of a black hole horizon, although precisely in this model one might also suspect the
converse, that black holes cannot form. If all string constants are kept positive, localized matter
configurations cannot exist, whereas all gravitational curvature must be associated with strings ---
there is no pure gravity in this model. So, if there is a black hole, strings will have to stick out
from it.

The absence of pure gravity degrees of freedom is intriguing. In a sense, matter here is ``unified" with
gravity, not, as in many models, because gravity generates particle-like degrees of freedom, but the
converse, because the matter degrees of freedom, here the string bits, carry around all the space-time
curvature there is.

It appears that one might have to decide also to allow for ``negative" strings, featuring surplus angles
rather than deficit angles. The question must be answered whether or not our newly formed string
segments can always be arranged such that they will all be positive ones. Judging from
Fig.~\ref{junctions.fig}$b$, this is unlikely but perhaps not impossible.

Also an important question is how to describe our \emph{choice} for a point in parameter space at
every intersection. Parameter space is compact, but the space of all possible collisions is not.
After accounting for all symmetries such as the Poincar\'e group at the center of mass, we are left
with a non-compact 4 dimensional space of all possible collision parameters. We need an infinite
dictionary to describe the parameters for what happens at all these possible interactions. As we
had to discover, this space is too large to exclude the existence of corners where further
complications arise.

Apart from all such questions, the model described here might be quite useful to address all sorts
of conceptual questions in classical and quantum gravity. The one thing it does not suffer from is
ultra-violet divergences, although the infrared question (the question as to what happens at large
distances and time intervals) will be quite difficult. Strings could terminate in infinitely dense
fractals of string segments, where they could close the universe.

\section*{Acknowledgements} The author thanks K.~Sfetsos for a discussion of this work. \def\RR{\hbox{Re\,}}
\def\II{\hbox{Im\,}} \appendix \newsecl{The algorithm for a quadrangle configuration}{app}
    In Figure \ref{strands.fig}$b$, we define the
holonomies of the external lines to be \(Q_{1,2,3,4}\), obeying
 \be Q_1\,Q_2\,Q_3\,Q_4=\mathbb{I}\ .\eel{qonefour} The internal lines have \(Q_{a,b,c,d}\) with
 \be Q_a=Q_d\,Q_1\ ,\quad Q_b=Q_a\,Q_2\ ,\quad Q_c=Q_b\,Q_3\ ,\quad Q_d=Q_c\,Q_4\ .\eel{intholon}
To do the calculations, we avoid square roots by setting
 \be Q_2=\W_2\ ,\qquad Q_1=Q_v\,\W_1\, Q_v^{-1}\ ,\qquad \W_i=\pmatrix{{1+it_i\over 1-it_i}&0\cr
 0&{1-it_i\over 1+it_i}} , \eel{qonetwo} where
 \be Q_v={1\over (1-v^2)(1+w^2)}\pmatrix{1+v^2&2v\cr 2v&1+v^2}\pmatrix{1-w^2&2wi\cr 2wi&1-w^2}\  .
 \eel{defcoeff}
Furthermore, \be Q_3=Q_1\,Q_2^{-1}Q_1^{-1}\ ,\qquad Q_4=Q_1Q_2\,Q_1^{-1}\,Q_2^{-1}Q_1^{-1}\ .
\eel{qthreefour} All \(Q\)'s obey \be \hbox{Im}\,\Tr(Q)=0\ ,\qquad |\hbox{Re}\,\Tr(Q)|<2\ . \eel{tracq}

The internal strings are then parameterized as follows:
 \be Q_a=\l\pmatrix{1+iy_{11}&x_{12}+iy_{12}\cr -\m_2(x_{12}+iy_{12})&1-iy_{11}}\ , \eel{paramqa}
where \(\l\) will be adjusted such that \(\det(Q_a)=1\). Tohether with Eq.~\eqn{intholon}, this
specifies all string parameters. We now define the holonomies \(Q_{e,f,g,h}\) as being the original
internal holonomies \(Q_{a,b,c,d}\) in the basis where \(Q_{1,2,3,4}\) is diagonal. This implies
 \be Q_e=Q_v^{-1}\,Q_a\,Q_v&,&\quad Q_f=Q_b\ \ , \nn\cr
 Q_g=Q_1^{-1}\,Q_c\,Q_1&,&\quad G_h=(Q_1Q_2Q_v)^{-1}Q_d(Q_1Q_2Q_v)\ . \eel{newbasis}
In this basis, they should all take the form \eqn{paramqa}, with the associated parameters \(\m_i\).
Therefore, we define the functions \(F_{ij}\) as follows:
 \be F_{i1}=\RR({Q_{i\,}}_1^1-{Q_{i\,}}_2^2)&,&\quad F_{i2}=\II({Q_{i\,}}_1^1+{Q_{i\,}}_2^2)\ ,\nn\cr
    F_{i3}=\RR(\m_i\,{Q_{i\,}}_2^1+{Q_{i\,}}_1^2)&,&\quad F_{i4}=\II({\m_i\,Q_{i\,}}_2^1-{Q_{i\,}}_1^2)\ . \eel{s1a-d}
If we choose the \(Q_i\) here to be \(Q_{e,f,g,h}\), these functions should all be zero.

When handling the most general case, the resulting expressions tend to become lengthy. It is more
illuminating to take an arbitrary example. We took:
 \be t_1=\fract13\ ,\quad t_2=\fract12\ ,\quad v=\fract17\ ,\quad w=\fract15\ . \eel{beginvw}
With this, the condition \(F_{41}=0\) yields
 \be y_{11}={1\over 148988448}(-3\,x_{12}(-39692619 + 46224949\,\m_2)+ \nn\cr
 28(-3849888 +y_{12}(33449 + 2251121\,\m_2)))\ . \eel{y11}
Then, \(F_{42}=0\) leads to
 \be y_{12}=-{15\,(179661440 + x_{12}(-75340697 + 354247143\,\m_2))\over 4\,(-902444771  +
 502645021\,\m_2)}\ . \eel{y12}
Next came the surprise: requiring \(F_{43}=F_{44}=0\) does \emph{not} fix the value of \(x_{12}\), but
in stead the value of \(\m_4\):
 \be\m_4={224726999641 + 40360716889\,\m_2\over 133691328551 - 379277148121\,\m_2}\ . \eel{mu4}
Indeed, with this value for \(\m_4\), the value of \(x_{12}\) is kept free. As a check, we find that,
with Eqs~\eqn{y11} and \eqn{y12}, all internal holonomies obey the string equation \(\II
\Tr(Q_{a,b,c,d}) =0\). Instead of \(x_{12}\), we could choose now \(\m_1\) as a new parameter.
Therefore, we check the functions \(F_{1i}\). Of these, \(F_{11}\) and \(F_{12}\) are already zero. Both
equations \(F_{13}=0\) and \(F_{14}=0\) lead to the same expression
 \be x_{12}= {-931364904960(-1 + \m_1)\over 1458158337341 - 2431046249155\,\m_2 +
 17\,\m_1(39825058285 + 73046360557\,\m_2)}\ .\nn\linebreak \eel{x12} This leaves the functions \(F_{3i}\) to be
checked. Again, \(F_{31}\) and \(F_{32}\) are already obeyed. The remaining two both give the same
result:
 \be\m_3  = {379277148121 + 40360716889\,\m_1\over 133691328551 - 224726999641\,\m_1}\ . \eel{mu3}
Since, in this case, both Eqs~\eqn{mu4} and \eqn{mu3} have a minus sign in their denominators, it is
easy to find positive values for \(\m_1\) and \(\m_2\) such that both \(\m_3\) and \(\m_4\) are positive
as well:
 \be  \m_1=\fract12\ ,\quad\m_2=\fract14\ ,\quad\m_3={266305004377\over 14218552487}\ ,
 \quad\m_4={939268715453\over 155488166083}\ . \eel{muval} The newly opened strings indeed also obey
Eq.~\eqn{tracq}:\pagebreak[0]
 \be \Tr(Q_a)=1.68394\ ,\quad \Tr(Q_b)=1.58638\ ,\quad  \Tr(Q_c)=1.88194\ ,\quad \Tr(Q_d)=1.37714\ ,
 \nn\linebreak
\eel{checkretr} where the values were rounded for clarity.

This good behavior, however, is due to the fact that the scattering is at high angles and
non-relativistic. If we do the same calculation for slightly different values:
 \be  t_1=\fract13\ ,\quad t_2=\fract12\ ,\quad v=\fract47\ ,\quad w=\fract45\ , \eel{beginvw2}
we get as our two equations:
 \be \m_3&=&-{23654969982136936 +  20734925253590287\,\m_1\over 19234905848692273 + 16644373216740712\,\m_1}\ , \nn\cr
 \m_4&=&-{16644373216740712 + 20734925253590287\,\m_2 \over 19234905848692273 + 23654969982136936\,\m_2}\ ;
 \eel{negativemu}
here, we see two equations that both are incompatible with positive values for \(\m_1\), \(\m_3\),
\(\m_2\) and \(\m_4\). As stated earlier, the general expressions for all values of \(t_1\), \(t_2\),
\(v\) and \(w\) in the allowed regions are too lengthy to be revealing.

Notice, finally, that the coefficients \(A\), \(B\), \(C\) and \(D\) for the two diagonals are clearly
related. This is due to the symmetry of the problem:
 \be \hbox{If}\quad \m_1={1\over \m_4}\quad\hbox{then}\quad\m_3={1\over\m_2}\ . \eel{murel} This
symmetry is due to the fact that our initial configuration was one with free strings approaching one another.
In this case, the existence of positive solutions for \(\m_1\) and \(\m_3\) automatically guarantees the
existence of positive solutions for \(\m_2\) and \(\m_4\), and \emph{vice versa}.

\end{document}